\documentclass[fleqn,usenatbib]{mnras}

\usepackage{newtxtext,newtxmath}

\usepackage[T1]{fontenc}
\usepackage{amsmath}

\usepackage{xcolor}
\usepackage{float}
\usepackage{hyperref}
\usepackage{upgreek}
\newcommand{\dmunits}{$\mathrm{pc\,cm^{-3}}$}
\usepackage{rotating}
\newcommand{\secref}[1]{\S\ref{#1}}
\usepackage{cleveref}
\usepackage{rotating}
\usepackage{graphicx}
\usepackage{caption}
\usepackage{subfig}
\usepackage{bm}
\usepackage{svg}
\usepackage{placeins}
\newcommand{\orcid}[1]{\href{https://orcid.org/#1}}
\crefformat{section}{\S#2#1#3} 
\crefformat{subsection}{\S#2#1#3}
\crefformat{subsubsection}{\S#2#1#3}
\DeclareRobustCommand{\VAN}[3]{#2}
\let\VANthebibliography\thebibliography
\def\thebibliography{\DeclareRobustCommand{\VAN}[3]{##3}\VANthebibliography}

\usepackage{graphicx}	
\usepackage{amsmath}	

\title{The second set of pulsar discoveries by CHIME/FRB/Pulsar: 14 Rotating Radio Transients and 7 pulsars}

\author[Dong et al.]{
\orcid{0000-0003-4098-5222}Fengqiu Adam Dong,$^{1}$\thanks{fengqiu.dong@gmail.com}
\orcid{0000-0002-1529-5169}Kathryn Crowter,$^{1}$ 
\orcid{0000-0001-8845-1225}Bradley W. Meyers,$^{1,2}$ 
\orcid{0000-0002-4795-697X}Ziggy Pleunis,$^{3}$
\orcid{0000-0001-9784-8670}Ingrid Stairs,$^{1}$\newauthor
\orcid{0000-0001-7509-0117}Chia Min Tan,$^{5,6}$
Tinyau Timothy Yu,$^{1}$
\orcid{0000-0001-8537-9299}Patrick J. Boyle,$^{5,6}$ 
\orcid{0000-0001-6422-8125}Amanda M. Cook,$^{3,4}$
\orcid{0000-0001-8384-5049}Emmanuel Fonseca,$^{7,8}$\newauthor
\orcid{0000-0002-3382-9558}B. M. Gaensler,$^{3,4}$
\orcid{0000-0003-1884-348X}Deborah C. Good $^{9,10}$
\orcid{0000-0001-9345-0307}Victoria Kaspi,$^{5,6}$
\orcid{0000-0002-2885-8485}James W. McKee,$^{11,12}$
\orcid{0000-0003-3367-1073}Chitrang Patel,$^{5,6}$\newauthor
\orcid{0000-0002-8912-0732}Aaron B. Pearlman$^{5,6}$
\\
$^{1}$Department of Physics and Astronomy, University of British Columbia, 6224 Agricultural Road, Vancouver, BC V6T 1Z1 Canada\\
$^{2}$International Centre for Radio Astronomy Research, Curtin University, Bentley, WA 6102, Australia\\
$^{3}$Dunlap Institute for Astronomy \& Astrophysics, University of Toronto, 50 St.~George Street, Toronto, ON M5S 3H4, Canada\\
$^{4}$David A. Dunlap Institute Department of Astronomy \& Astrophysics, University of Toronto, 50 St. George Street, Toronto, Ontario, Canada M5S 3H4\\
$^{5}$Department of Physics, McGill University, 3600 rue University, Montr\'eal, QC H3A 2T8, Canada\\
$^{6}$McGill Space Institute, McGill University, 3550 rue University, Montr\'eal, QC H3A 2A7, Canada\\
$^{7}$Department of Physics and Astronomy, West Virginia University, Morgantown, WV 26506-6315, USA\\
$^{8}$Center for Gravitational Waves and Cosmology, Chestnut Ridge Research Building, Morgantown, WV 26505, USA\\
$^{9}$Department of Physics, University of Connecticut, 196A Auditorium Rd Unit 3046, Storrs, CT, USA; Center for Computational Astrophysics,\\
$^{10}$Flatiron Institute, 162 5th Avenue, New York, New York, 10010, USA\\
 $^{11}$E.A. Milne Centre for Astrophysics, University of Hull, Cottingham Road, Kingston-upon-Hull, HU6 7RX, UK\\
 $^{12}$Centre of Excellence for Data Science, Artificial Intelligence and Modelling (DAIM), University of Hull, Cottingham Road, Kingston-upon-Hull, HU6 7RX, UK
}

\date{Accepted 2023 June 29 Revised 2023 June 29 Submitted 2022 October 17}

\pubyear{2015}

\begin{document}
\label{firstpage}
\pagerange{\pageref{firstpage}--\pageref{lastpage}}
\maketitle

\begin{abstract}
The Canadian Hydrogen Mapping Experiment (CHIME) is a radio telescope located in British Columbia, Canada. The large FOV allows CHIME/FRB to be an exceptional pulsar and Rotating Radio Transient (RRAT) finding machine, despite saving only the metadata of incoming Galactic events. We have developed a pipeline to search for pulsar/RRAT candidates using DBSCAN, a clustering algorithm. Follow-up observations are then scheduled with the more sensitive CHIME/Pulsar instrument capable of near-daily high time resolution spectra observations. We have developed the CHIME/Pulsar Single Pulse Pipeline to automate the processing of CHIME/Pulsar search-mode data. We report the discovery of 21 new Galactic sources, with 14 RRATs, 6 isolated long-period pulsars and 1 binary system. Owing to CHIME/Pulsar's observations we have obtained timing solutions for 8 of the 14 RRATs along with all the regular pulsars and the binary system. Notably we report that the binary system is in a long orbit of 412 days with a minimum companion mass of 0.1303 solar masses and no evidence of an optical companion within 10" of the pulsar position. This highlights that working synergistically with CHIME/FRB's large survey volume CHIME/Pulsar can obtain arc second localisations for low burst rate RRATs though pulsar timing. We find that the properties of our newly discovered RRATs are consistent with those of the presently known population. They tend to have lower burst rates than those found in previous surveys, which is likely due to survey bias rather than the underlying population.

\end{abstract}

\begin{keywords}
pulsars: individual:  -- stars: neutron -- pulsars: general
\end{keywords}


\section{Introduction} \label{sec:intro}
Pulsars act as lighthouses in radio frequencies with beams that sweep over our observatories at regular intervals. They are remnants of massive stars that have undergone core collapse \citep{Gold1968}; the remains are a degenerate core at supranuclear densities. Furthermore, pulsar emissions are observed to be periodic, thus, timing models can precisely describe their spin evolution \citep{Hobbs2004,Lynch_2018}. Recently, subclasses of neutron stars have provoked fervent interest in the field of transient astronomy \citep{Kaspi2010}. One subclass discovered in 2006 \citep{McLaughlin2006,Keane2011} is the Rotating Radio Transients (RRATs); they are radio-emitting neutron stars that by definition can only be detected in single pulses as opposed to a Fourier domain search \citep{Keane2011,Burke-Spolaor2012}. RRATs have garnered more attention recently as single pulse searches have grown in popularity. They are particularly intriguing as the RRATs are thought to be as populous as conventional radio pulsars \citep{McLaughlin2006,Keane2011} and can be informative of the total Galactic distribution of neutron stars. However, while many surveys have been conducted, only 136 RRATs have been found\footnote{http://astro.phys.wvu.edu/rratalog/}. The incomplete knowledge of the nulling fraction of RRATs makes population studies more difficult than conventional pulsars. Another interesting feature of RRATs is their similar radio presentations to Fast Radio Bursts (FRBs) which are extremely bright, millisecond timescale extragalactic bursts. There have even been attempts to find FRBs in mislabelled RRATs \citep{Keane2016_2,Rane2017}. Another subclass of neutron stars are magnetars, which are thought to be young, highly magnetised neutron stars \citep{Kaspi2017} and may explain high energy phenomena such as FRBs \citep{TheCHIMECollaboration2020,Bochenek2020}. Therefore, finding more RRATs and magnetars can inform astronomers about the population of Galactic pulsars, pulsar distributions and can potentially disentangle the population of RRATs, magnetars and FRBs.

While there are over 3300 known pulsars to date \citep{Manchester2005}\footnote{https://www.atnf.csiro.au/research/pulsar/psrcat/ version 1.7} only $\sim$4\% of them are RRATs. The low proportion is exacerbated by the lack of knowledge of the nulling fraction of RRATs. It is possible that large surveys like the Parkes Multibeam Survey \citep{Manchester2001} only captured a fraction of the total RRATs \citep{Mickaliger2018} encompassed in the survey sky area, resulting in an incomplete picture of the pulsar population. This can be remedied as next-generation and Square Kilometer Array precursor telescopes begin large-scale pulsar and fast-transient searches \citep{Johnston2008,Tzioumis2010,Coenen2014,Swainston_2021}.
Many of the telescopes contributing to RRAT discoveries have unique characteristics such as wide fields of view e.g. CHIME \citep{CHIME_FRB_2018,CHIME_Pulsar_2020}, LOFAR \citep{lofar2013}, MWA \citep{Wayth2018,Tingay2013}, ASKAP \citep{Johnston2007,Macquart2010} or outstanding sensitivity e.g. FAST \citep{Qian2019}. This leads to rapid coverage of the celestial sphere. The large survey volumes necessitate the development of automated pipelines and algorithms to identify pulsar and RRAT candidates.\\

Pulsar and RRAT surveys are conventionally split into two categories, untargeted and targeted. Untargeted surveys will search a grid of celestial positions \citep[e.g.,][]{Manchester2001,Parent_2022,Parent2019,Stovall_2014} while targeted surveys will have deep exposures on particularly interesting sky locations \citep[e.g.,][]{Pan_2021,Manchester1991,Rifolfi2021,Yan_2021}. In this study, we use the untargeted CHIME telescope's FRB instrument (CHIME/FRB), which has fixed beams on the transiting sky. The CHIME/FRB survey is especially unique due to the $\sim 200$ square degree field of view on the transiting sky, allowing us a daily search grid of the down to a declination of -10$^\circ$. \\

The primary difference in detecting RRATs and FRBs is the Galactic nature of RRATs, quantified by their dispersion measure (DM), a measure of the time delay in the signal because of the interstellar and intergalactic medium. The DM follows a $f^{-2}$ law, where $f$ if observing frequency, such that lower frequencies are delayed longer. FRBs tend to have very high DMs due to their extragalactic nature while RRATs tend to have lower DMs. Thus, as most search pipelines first dedisperses incoming data, the two types of radio transients are often discovered in similar fashions \citep{Caleb2016,Keane2018,Bhandari2018,TerVeen2019}.\\
 
The rapid survey capability of CHIME/FRB yields has yielded $\sim 1.5\times 10^8$ single pulse events; many of them are false positives. For each event, numerous statistics are generated; some useful ones are described in \secref{sec:CHIME/FRB}. Crucially, we record the celestial locations in J2000 coordinates as Right Ascension (RA) and Declination (Dec) along with the DM. In this study, we have developed a search pipeline called the CHIME/FRB Metadata Clustering Analysis (CHIMEMCA) to filter and cluster single pulse events to produce pulsar and RRAT candidates. CHIMEMCA uses DBSCAN \citep{Ester96adensity-based}, an unsupervised machine learning algorithm, to cluster events in RA, Dec and DM. We follow up promising CHIMEMCA candidates with the higher sensitivity of the CHIME/Pulsar system (\secref{sec:CHIME/Pulsar}, \citet{CHIME_Pulsar_2020}). CHIME/Pulsar is a radio instrument that allows for monitoring of 10 sources at a time with steerable digital tracking beams. Each tracking beam is able to produce high time resolution spectra (search-mode hereafter). This allows almost daily coverage of the CHIMEMCA pulsar and RRAT candidates. The high observation cadence drove the development of a second automated pipeline, the CHIME/Pulsar Single-pulse PIPEline (CHIPSPIPE). CHIPSPIPE is a PRESTO \citep{2001PhDT.......123R,2011ascl.soft07017R} based pipeline that also utilises SPEGID \citep{Pang2018} and FETCH \citep{Agarwal2020} to process the CHIME/Pulsar search-mode data stream. In this work, we present details on 14 new RRATS and 7 new pulsars discovered using CHIMEMCA and CHIPSPIPE.\\

The outline for this paper is as follows:
\begin{itemize}

\item \secref{sec:CHIME} Covers the capabilities of the various backends of the CHIME telescope, namely, CHIME/Pulsar and CHIME/FRB.

\item \secref{sec:CHIMEMCA} Introduces the unsupervised machine learning pipeline (CHIMEMCA) used to find new Galactic source candidates.

\item \secref{sec:CHIPSPIPE} Introduces the automated search-mode data reduction pipeline for CHIME/Pulsar (CHIPSPIPE).

\item \secref{sec:reults} Presents the timing analysis and single pulse distributions of 21 new Galactic pulsars and RRATs.

\item \secref{sec:discussion} Discusses of the implications of all the CHIME/FRB pulsar and RRAT discoveries and compares them with previously known populations.
 
\end{itemize}

\section{CHIME Telescope}\label{sec:CHIME}

\subsection{CHIME/FRB}\label{sec:CHIME/FRB}
CHIME/FRB is a backend instrument to the CHIME telescope specifically designed to detect FRBs. CHIME/FRB receives 1024 total intensity (Stokes I) data streams at a time resolution of 0.983 ms and a frequency resolution of 24.4KHz. Each stream represents one of the 1024 tied-array beams that are digitally formed in the CHIME main beam \citep{CHIME_FRB_2018}. Furthermore, each of the tied-array beams has a FWHM of $\sim$20’-40’. 
\\To accommodate the large data stream CHIME/FRB continuously runs a real-time pipeline that

\begin{enumerate}
    \item Dedisperses the data using the \texttt{BONSAI}\footnote{https://github.com/kmsmith137/rf\_pipelines} tree dedispersion algorithm
    \item Employs RFI mitigation based on running data statistics \citep{Rafiei-Ravandi2022}
    \item Groups all candidates such that a single event detected in multiple beams is detected as one event.
    \item Grades every event using a machine learning RFI classifier called \texttt{RFISifter}
    \item Assigns a flag to events that are spatially (including DM) close to nearby known pulsars. This is called the \texttt{KnownSourceSifter}
    \item Determines whether an incoming event has Galactic or extragalactic DM. This is done by using the YMW16 \citep{Yao2016} or the NE2001 \citep{Cordes2002} DM models. 
    \item Issues a callback (request to save data) for total intensity and raw voltage data if the event is extragalactic.
    \item Saves metadata for all events.
\end{enumerate}
CHIME/FRB actively monitors $\sim 200$ square degrees of the sky at all times, thus has a high chance of detecting sporadic single pulses from RRATs and pulsars. While the total intensity is not saved for Galactic events, the metadata provide a rich database to search for Galactic pulsars and RRATs. The metadata include:
\begin{itemize}
    \item Right Ascension 
    \item Declination 
    \item Dispersion Measure
    \item \texttt{RFISifter} grade : A grade between 1 (RFI) and 10 (astrophysical) that \texttt{RFISifter} has given an event.
    \item \texttt{KnownSourceSifter} associations : The association of an event to known Galactic pulsars or repeating FRBs.
    \item Beam activity : The number of synthesised beams that exceeded a fiducial Signal to Noise Ratio (SNR) threshold for a given event
\end{itemize}
We feed this information into a unsupervised machine learning algorithm to produce Galactic transient candidates. This is described in \secref{sec:CHIMEMCA}.

\subsection{CHIME/Pulsar}\label{sec:CHIME/Pulsar}
The CHIME/Pulsar backend can observe 10 sources at one time using digitally formed beams, and with continuous operation CHIME/Pulsar observes 400-500 known pulsars per day. The typical transit times lasts $\sim 10$ minutes, which is sufficient to enable pulsar timing and searching observations. Each of the instrument's 10 tied-array beams has a FWHM of $\sim 0.25^\circ$ at 800MHz. CHIME/Pulsar also operates in 3 capacities: fold, search and baseband modes. Fold-mode is for phase folding pulsar spectra with known ephemerides. search-mode records the high time resolution spectra of the total intensity and is the default for surveys and new candidates. We note that search-mode data can be folded at a later time if required. To increase sensitivity in these two modes, we can coherently dedisperse the data up to a DM of $\sim$500\dmunits. Baseband mode is reserved for high time resolution and polarisation studies and records the raw beam-formed data stream. A thorough discussion of the CHIME/Pulsar system can be found in \cite{CHIME_Pulsar_2020}.

In the this work, we use both search and fold-modes. search-mode has a time resolution of $327.68 \mu s$, greatly improving on CHIME/FRB detections and sensitivity. All new pulsar candidates start with search-mode observations. Once detected, if the source is a conventional radio pulsar and a period is identified, it will be switched to fold-mode observations.

\section{Chime/FRB Metadata Clustering Analysis}\label{sec:CHIMEMCA}
In this section we present a repeating astrophysical source search pipeline called the CHIME/FRB Metadata Clustering Analysis (CHIMEMCA). We use the CHIME/FRB metadatabase of Galactic events by clustering in RA, Dec and DM. As a result, many candidates are produced to be followed up with CHIME/Pulsar. Due to CHIME/FRB's sky coverage and sensitivity to single pulse events, most Galactic discoveries detected are RRATs that have likely been missed in previous surveys. We note that this analysis is equally adept at identifying extragalactic transients as well as Galactic pulsars, and has been used to identify repeating FRBs \citep{CHIME2023}.\\
\subsection{DBSCAN}\label{DBSCAN_CHIMEMCA}

To systematically search our metadata database for Galactic pulsars, we use an unsupervised machine learning algorithm called Density-based spatial clustering of applications with noise (DBSCAN) \citep{Ester96adensity-based}. DBSCAN is an algorithm to find clusters of points in a n-dimensional space. In our case, this is the RA ($\alpha$), Dec ($\delta$) and DM space. DBSCAN has two benefits over other methods such as K-means \citep{Macqueen67}. Firstly, it is able to easily trace arcs in RA-Dec space. The arcs occur because of erroneous localisations as bright sources move into the CHIME sidelobes. However, the sources that are known to do this need to be bright enough to overcome the attenuation of the CHIME sidelobes (e.g. the Crab Pulsar). Secondly, DBSCAN, unlike K-means, can form an unbounded number of clusters and therefore does not require an arbitrary choice in the number of clusters. We use DBSCAN from the \texttt{sklearn}\footnote{https://scikit-learn.org/} package in \texttt{python}. This version of DBSCAN requires three parameters: $\epsilon$, the minimum distance between cluster points, \texttt{min\_samples}, the minimum number of events per cluster and \texttt{metric}, the distance metric between events. Firstly, we choose $\epsilon$ to be a function of the uncertainties of each event. Due to CHIME’s design, the uncertainties in RA are not constant across our declination range. For all declinations, the uncertainty of RA goes as $\sigma_{\alpha}\propto\frac{1}{cos(\delta)}$, greatly inflating our uncertainties at declinations near the pole. We gave fiducial values for declination and DM errors, $\sigma_{\delta}=0.5\deg$, $\sigma_{DM}=13\text{pc\,cm}^{-3}$. These were chosen because of the full width half maximums of the beam and the largest dedispersion uncertainties from the tree dedispersion and peak-finding algorithm. Instead of varying $\epsilon$, we equivalently scale each parameter by their respective errors and then set $\epsilon=1$.\\
A second issue is that RA ($\alpha$) is cyclic at 360$ ^{\circ}$. To deal with this, we used a cyclic function to map RA to a value between -90 and 90 as described by equation \ref{eq:fn1} and \ref{eq:fn2}, replacing $\alpha$ with $\zeta_1(\alpha)$ and $\zeta_2(\alpha)$. This way, events with high RA will be clustered with those with low RA, assuming that the Dec and DM are also sufficiently close. 

\begin{equation}
\begin{split}
\zeta_1(\alpha) =
\begin{cases}
    \alpha & \alpha < 90\degr\\
     180\degr - \alpha & 90\degr \leq \alpha < 270\degr \\
     \alpha - 360\degr & 270\degr \leq \alpha < 360\degr\\
\end{cases}
\label{eq:fn1}%
\end{split}
\end{equation}

\begin{equation}
\begin{split}
\zeta_2(\alpha) =
\begin{cases}
     90\degr - \alpha & \alpha < 180\degr\\
     \alpha - 270\degr & 180\degr \leq\alpha \\
\end{cases}
\label{eq:fn2}%
\end{split}
\end{equation}
The errors for $\zeta_i$ are given by:
\begin{equation}
\bm{\sigma}_{\zeta_i}^2 = \left(\frac{\partial \zeta_i}{\partial \alpha}\right)^2 \bm{\sigma}_\alpha^2
\label{eq:error_prop}
\end{equation}
where $\sigma_{\zeta_i}$ is the error of the functions described in equations~(\ref{eq:fn1}--\ref{eq:fn2}) and $\sigma_\alpha$ is the error in RA.
The vector space fed into DBSCAN is
\begin{equation}
\mathbf{V}_k = \frac{\mathbf{v}_k}{\bm{\sigma}_k}
\label{eq:scale}
\end{equation}
where $\mathbf{v}_k=[\zeta_1, \zeta_2, \delta, {\rm DM}]$, $\bm{\sigma_k} = [\sigma_{\zeta_i}, \sigma_{\zeta_i}, \sigma_{\delta}, \sigma_{\rm DM}]$, and $k$ subscripts each individual event.\\
Finally, we chose \texttt{min\_points}=2 and \texttt{metric}=\texttt{Chebyshev}. The Chebyshev distance is defined as
\begin{equation}
    D_\mathrm{Chebyshev} =\text{max}_i(|x_i-y_i|)
\end{equation}
where $i$ is indexed over $\zeta_1$, $\zeta_2$, $\delta$ and DM and $x$ and $y$ are two separate events. In this case, as each parameter has been scaled by their respective errors $D_\mathrm{Chebyshev}$ denotes the distance between two events in $\sigma_i$ units. Thus $\epsilon$ is set to 1 such that any two events separated by less than $\sigma_i$ are clustered together.
As we have errors in RA, Dec and DM, a Chebyshev metric allows two events to be clustered even if they are on the limits of the allowed error budget of all clustering parameters. One can choose other metrics like Euclidean without much difference in the final results.

\begin{figure*}%
    \centering
    \subfloat[\centering]{{\includegraphics[width=0.47\textwidth]{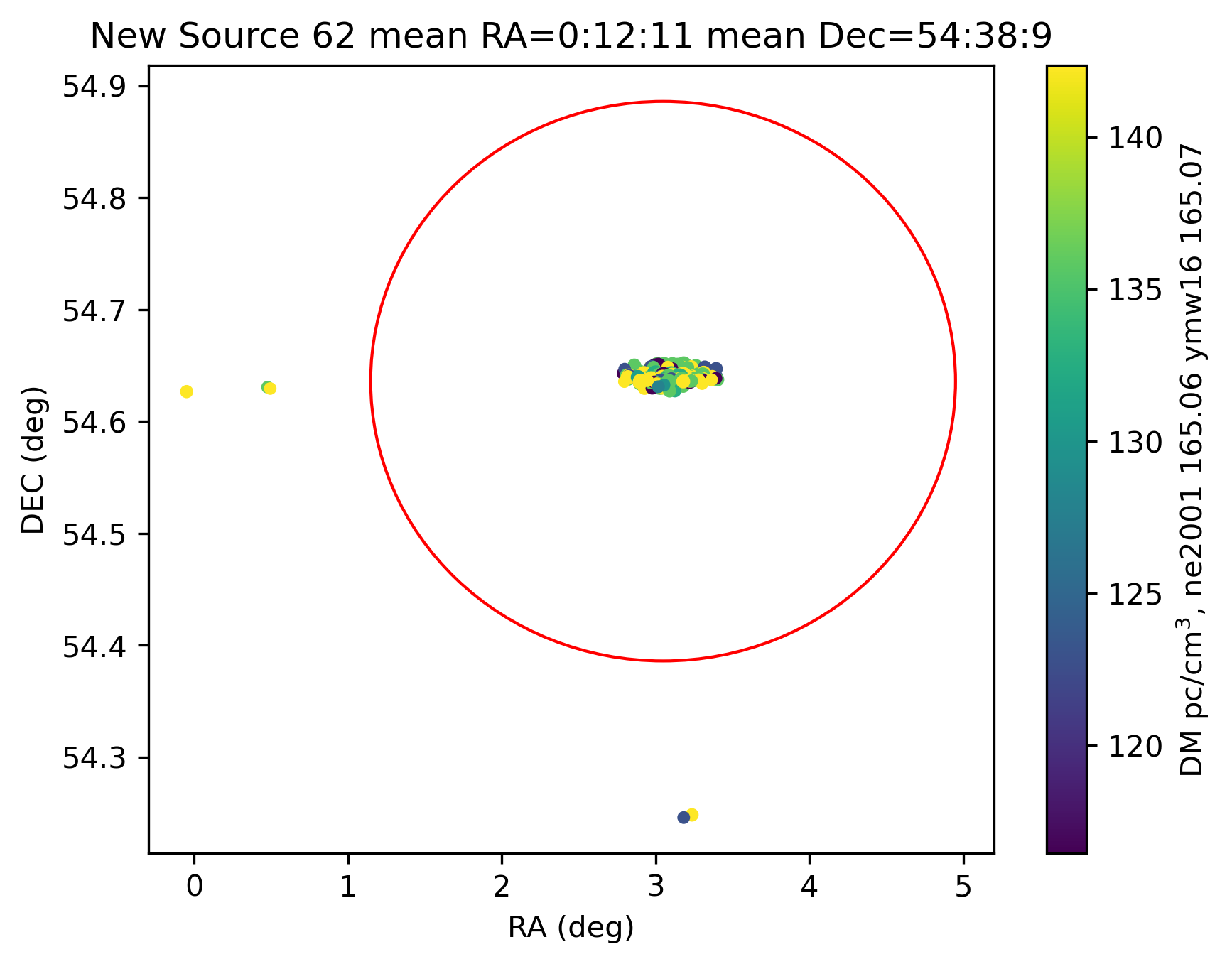} }}%
    \qquad
    \subfloat[\centering]{{\includegraphics[width=0.47\textwidth]{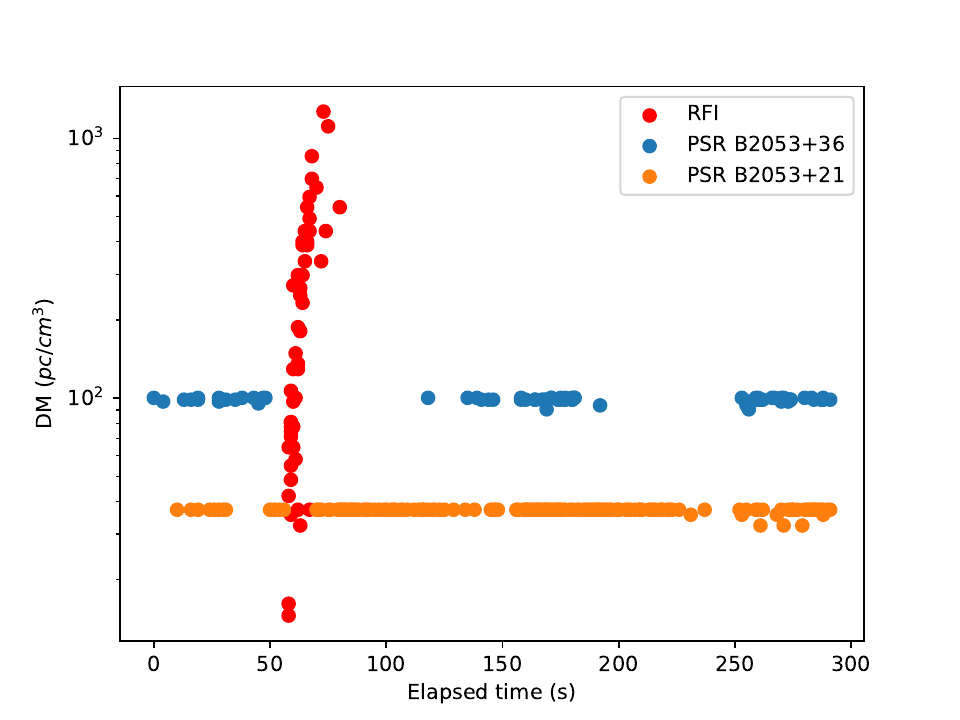} }}%
    \caption{(a) Cluster number 62 (PSR J0012+5431). We find that PSR J0012+5431's DM is $\sim20\%$ below the Galactic DM using both the YMW16 \citep{Yao2016} and NE2001 \citep{Cordes2002} models. The red ellipse gives a scale of CHIME/FRB beam size at this declination. The points outside the red circle are detections of PSR J0012+5431 outside the FWHM or are detections by a neighbouring synthesized beam. The axis label for the colour bar indicates the largest DM ($\text{pc\,cm}^{-3}$) in this direction according to the NE2001 and YMW16 models. (b) The DM-time plot is used to identify RFI; the horizontal blue and orange lines show pulsar transits while the red stripe shows an RFI streak.}%
    \label{fig:Clustering}%
\end{figure*}

\subsection{Pipeline} \label{DBSCAN}
For this work, we queried the metadata database from 2018-10-01 to 2021-11-15. This set of data is in excess of $1.5\times10^8$ events, so RFI and known source filtering are vital for truly unknown astrophysical candidates.\\
For RFI filtering; we take a variety of cuts based on the statistics in the metadata. First we take only data with SNR$>$9; this ensures that only the brightest bursts remain. Following that, in line with CHIME/FRB's main pipeline, we cut on \texttt{RFISifter} grade$>$5 (\secref{sec:CHIME/FRB}). Finally we cut on beam activation (\secref{sec:CHIME/FRB}), whereby events with \texttt{beam\_activation}$>$50 are classified as RFI. We set this criterion on beam activation because we find that true astrophysical pulses are not typically detected in more than 50 beams simultaneously. Note that this should not hurt the completeness of the survey significantly, as the single pulse luminosities of pulsars are thought to follow log-normal distributions \citep{Burke-Spolaor2011,Mickaliger2018,Meyers2018,Meyers2019} thus any astrophysical events that are detected in more than 50 CHIME/FRB beams should be extremely rare. The final filtering step is to remove any known Galactic pulsars via events flagged by the \texttt{KnownSourceSifter} (\secref{sec:CHIME/FRB}).\\
Following the RFI and known source filtering process, we apply the technique described in \secref{DBSCAN_CHIMEMCA} to produce clusters of events that eventually become pulsar candidates. A typical cluster identified by DBSCAN is shown in Figure \ref{fig:Clustering} (a) which depicts PSR J0012+5431, a RRAT discovered using CHIMEMCA and presented in this work. Properties such as high positional clumping within the CHIME/FRB beam size and low DM variation are indicative of a new Galactic source. There are points outside of the FWHM circle; these are likely detections of PSR J0012+5431 in a neighbouring beam or sidelobe.

To ensure that RFI has not circumvented the cuts and is not polluting the candidates table, each cluster is inspected by eye. As each cluster comprises many events, to check for RFI, we take each event and make a DM-time plot of all other temporally adjacent events. RFI manifests as near vertical lines while pulsars and FRBs are isolated points or horizontal lines. In Figure \ref{fig:Clustering} (b) there is a RFI streak at $\sim$ 60s whereas all the points on the horizontal lines belong to either PSR B2053+36 or PSR B2053+21.
\\After CHIMEMCA has identified a new candidate, a secondary program checks the average cluster positions against two pulsar survey "scrapers"  \footnote{http://hosting.astro.cornell.edu/\textasciitilde deneva/tabscr/tabscr.php}\textsuperscript{,}\footnote{https://pulsar.cgca-hub.org/}. This is a check against pulsars discovered by other surveys that have not yet been added into the \texttt{KnownSourceSifter} database. The clusters that survive this vetting process are pulsar/RRAT candidates and assigned observation time with CHIME/Pulsar.\\
While the majority of the discoveries presented in this work has been found via CHIMEMCA, one pulsar and one RRAT were found serendipitously by CHIME/FRB operators. These two sources were also identified by CHIMEMCA subsequently.
\subsection{CHIME/FRB detection}
In addition to candidate searching, we can also perform long timescale activity analyses of RRATs using the CHIME/FRB metadata. This involves finding all events for a source with SNR greater than 8. Following that, we show the daily sensitivity as characterised by the detection of known pulsars and the exposures for the source's sky location. The same methods were used as \citet{Good_2021}. All plots are hosted on https://www.chime-frb.ca/galactic. 

\section{CHIME/Pulsar Observations and the CHIME/Pulsar Single-pulse Pipeline}\label{sec:CHIPSPIPE}
 CHIME/Pulsar observes on a probability based schedule, depending on the current observing load \citep{CHIME_Pulsar_2020}. This ensures that new candidates can be observed even in over-subscribed sky regions. One benefit of observing with CHIME/Pulsar is that it has higher sensitivity than CHIME/FRB due to coherent dedispersion, greater time resolution and the ability to track the source over the transit. Furthermore, as we do not collect intensity data from CHIME/FRB we require a detection from CHIME/Pulsar to claim a new pulsar discovery. A complete list of published and unpublished pulsars is available on the CHIME/FRB Galactic sources webpage \footnote{https://www.chime-frb.ca/galactic}. 

With CHIME/Pulsar, we always initially take search-mode data for every pulsar candidate. This is quickly changed to fold-mode after a candidate is confirmed to be a pulsar and the period is determined; otherwise, for example for RRATs, we continue to take search-mode observations.
\subsection{CHIME/Pulsar data reduction and pipeline}
\begin{figure}
    \centering
    \includegraphics[width=0.5\textwidth]{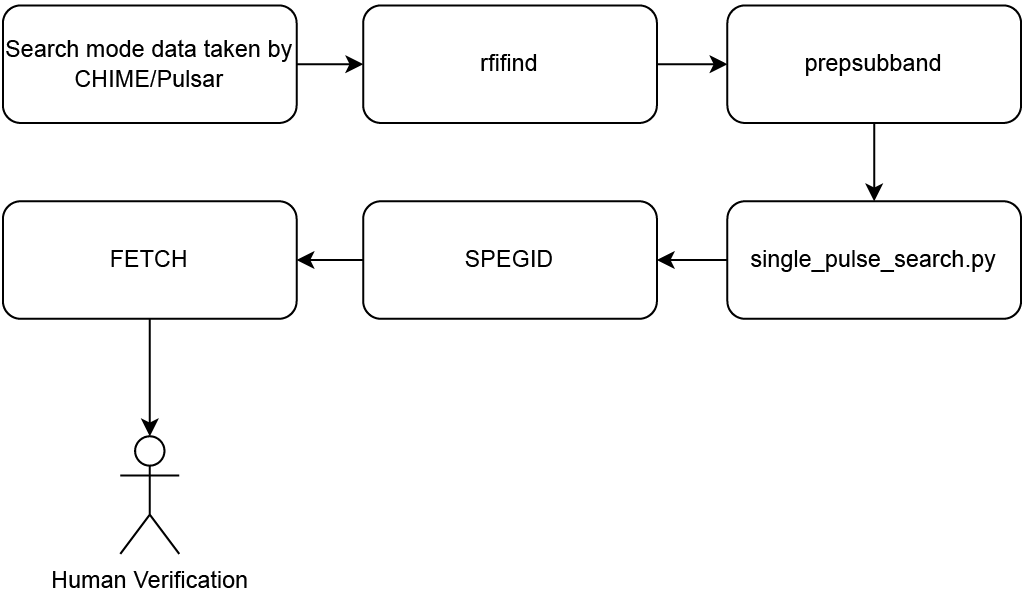}
    \caption{Block diagram for CHIPSPIPE. The majority of the tools used belong to the \texttt{PRESTO} software suite, however, we use \texttt{SPEGID} to generate single pulse candidates and \texttt{FETCH} to grade them on their astrophysical nature.}
    \label{fig:automated_filterbank}
\end{figure}

Depending on the observing pressure at a particular sky location, CHIME/Pulsar can achieve an almost daily cadence for observing pulsar/RRAT candidates. Although some new candidates fall in oversubscribed sky regions leading to a reduction in observations, this still yields 2466 separate search-mode pointings. Evaluating and processing extensive datasets quickly become infeasible for a human. To automatically handle the data sets, we have developed the open sourced CHIME/Pulsar Single-pulse PIPEline (CHIPSPIPE) \footnote{https://github.com/CHIME-Pulsar-Timing/CHIME-Pulsar\_automated\_filterbank}, a PRESTO based search pipeline that also incorporates \texttt{SPEGID} \citep{Pang2018} and \texttt{FETCH} \citep{Agarwal2020}. We use "model \texttt{a}" from \texttt{FETCH} as it was reported to have the highest Fscore by \cite{Agarwal2020} and we are not aware of any limitations to this model. A block diagram of CHIPSPIPE is shown in \Cref{fig:automated_filterbank}. In addition to CHIPSPIPE, we use \texttt{riptide} \citep{Morello2020}, a Fast Folding Algorithm (FFA) and \texttt{prepfold}, a \texttt{PRESTO} module, to obtain the folded profiles of pulsars. For RRAT data that cannot be folded, we use \texttt{rrat\_period} or \texttt{rrat\_period\_multiday} included in the \texttt{PRESTO} package to estimate the periods.\\
To handle RFI in the search-mode data, we use the \texttt{rfifind} software included in \texttt{PRESTO} to excise contaminated data. \texttt{rfifind} is able to remove transient narrow band as well as periodic RFI. We also apply a static mask that will remove most of the known bad channels at CHIME. This process reduces our band by $\sim 30\%$. Finally, a known issue is dropped packets in the data streaming process when the search-mode observation is recorded. A dropped packet results in a near 0 value in the search-mode data. We remove this by looking for drop-outs below 4.5$\sigma$ of the median for each frequency channel and replacing them with the median.\\
When taking data, each of the 1024 frequency channels is coherently dedispersed to the nominal DM of the candidate.  The \texttt{PRESTO} software is then used to search a small DM range of  $\pm 20 \text{pc\,cm}^{-3}$ around the cluster DM.  The \texttt{DDplan.py} script prepares a list of trial DMs, and the \texttt{prepsubband} program is used to incoherently dedisperse the frequency channels and generate a time series for each trial DM. As we do not need to search the full DM range, the processing time is reduced drastically, especially for high DM candidates. \\

Each time series generated by \texttt{prepsubband} is searched using the \texttt{single\_pulse\_search.py} software included in \texttt{PRESTO}. This generates a series of text files containing single pulse time of arrivals (TOAs), SNRs and searched boxcar widths.\\
As the output for \texttt{single\_pulse\_search.py} is often time consuming for a human to check we use \texttt{SPEGID}, a DBSCAN based clustering algorithm, to find single burst candidates \citep{Pang2018}. SPEGID will rank candidates from 1 (most likely astrophysical) to 5 (least likely astrophysical). The input to \texttt{SPEGID} is the text files output by \texttt{single\_pulse\_search.py}. \texttt{SPEGID} then makes an assessment on whether a series of single pulses is characteristic of an astrophysical or RFI signal, thereby generating a list of pulse candidates.\\ 
\texttt{SPEGID} candidates that receive a rank of 1 or 2 (likely astrophysical) are passed to \texttt{FETCH}, a deep-learning based classifier used to detect impulsive bursts from \texttt{sigproc}-filterbank data \citep{Agarwal2020}. Finally, all astrophysical bursts are extracted for human review.\\
FETCH generates diagnostic waterfall plots as well as a corresponding grade. A human then needs to review all positively graded bursts from FETCH and save those that are truly astrophysical. In this process the human reviewer only keeps FETCH results with SNR$>$6.\\
For period extraction, \texttt{rrat\_period} and \texttt{rrat\_period\_multiday} are used with the timestamps of pulses that survive human vetting. Both packages are included in \texttt{PRESTO} and operate on finding the largest common factor for time in between closely spaced pulses from a single source. If a period results from this, \texttt{prepfold} is used to fold the profile and, in the case of regular pulsars, refine the period. If either \texttt{prepfold} or \texttt{rrat\_period} are unsuccessful, \texttt{riptide} is then used to perform a wider time-domain search than is possible with \texttt{prepfold} while using relatively few computing resources.

\subsection{Timing and fold-mode} \label{sec:timing}
We observe pulsars via CHIME/Pulsar fold-mode observation. This mode uses an ephemeris generated by the initial search-mode detection. The folding of the data is done via two parallel methods:
\begin{enumerate}
    \item We use \texttt{PSRCHIVE} and the suite of tools it provides like \texttt{pazi} to manage the RFI and use \texttt{pam} to manage archives.
    \item We use \texttt{PRESTO's} \texttt{prepfold} to fold and refine the period derived from \texttt{rrat\_period}.
\end{enumerate}
For the sources that can be folded, the general procedure is described by \citet{Good_2021} where the Digital Signal Processing Software for Pulsar Astronomy \citep[\texttt{DSPSR}]{VanStraten2011} is used. RFI mitigation is done through \texttt{paz} and TOAs are generated via \texttt{pat} for all search-mode data. Finally, we determine the timing solution using \texttt{TEMPO}\footnote{http://tempo.sourceforge.net/} \citep{TEMPO} or \texttt{TEMPO2}\footnote{https://bitbucket.org/psrsoft/tempo2} \citep{Edwards2006,Hobbs2006}.

For the timing of the RRAT discoveries, we use \texttt{DSPSR} to extract and fold 30 seconds of data around the reported detection times from \texttt{SPEGID}. We then extract the sub-integration where the pulse is present, generate TOAs and determine the timing solutions from the single pulses based on the method above.

\section{Results and Discoveries} \label{sec:reults}
\begin{table*}
\caption{Pulsar observations table ordered by RA with RRATs followed by regular pulsars}
\centering
\begin{tabular}{|l|c|c|c|c|c|}
\hline\hline
RRAT/PSR      & Timing Solution Provided & \# Search-mode obs & \# Fold-mode obs &  Type & Observation Hours\\\hline\hline
PSR J0012+5431 & Yes    & 116             & -                          & RRAT & 46.33 \\
PSR J0653-06 & Yes     & 146             & -                           & RRAT & 33.92 \\
PSR J0741+17 & No     & 119             & -                         & RRAT & 28.73 \\
PSR J1105+02 & Yes    & 195             & -                           & RRAT & 44.18 \\
PSR J1130+0921 & Yes    & 319             & -                     & RRAT  & 71.08\\
PSR J1541+4703 & Yes    & 248             & -                           & RRAT & 79.05 \\
PSR J2008+3758 & Yes    & 65              & -                          & RRAT & 18.09 \\
PSR J2113+73 & No     & 73              & -                           & RRAT & 59.40 \\
PSR J2138+69 & No     & 84              & -                            & RRAT & 53.64 \\
PSR J2215+4524 & Yes    & 130             & -                        & RRAT & 41.11 \\
PSR J2237+2828 & Yes    & 272             & -                          & RRAT & 69.64 \\
PSR J2316+75 & No     & 77              & -                           & RRAT & 63.78 \\
PSR J2221+81 & No     & 91              & -                           & RRAT & 123.03 \\
PSR J2355+1523 & Yes    & 531             & -                          & RRAT & 101.72 \\\hline
PSR J0227+3356 & Yes    & -               & 320                         & Pulsar  & 83.43 \\
PSR J0658+2936 & Yes    & -               & 170                         & Pulsar  & 46.04 \\
PSR J1048+5349 & Yes    & 241             & -                          & Pulsar  & 92.30 \\
PSR J1943+5815 & Yes    & -               & 32                  & Pulsar   & 13.84\\
PSR J2057+4701 & Yes    & -               & 142                         & Pulsar  & 47.60 \\
PSR J2116+3701 & Yes    & -               & 247                         & Pulsar  &  72.64\\
PSR J2208+4610 & Yes    & -               & 183                         & Pulsar  & 61.05 \\
\hline
\end{tabular}
\label{tab:all_psr}
\end{table*}
We present the discovery of 21 new Galactic pulsars and RRATs with timing solutions for 15 of them. Additionally, more than half of the discovered sources (14) were RRATs and we have successfully obtained timing solutions for 8 of them. This showcases CHIME/Pulsar's capabilities at timing intermittent sources due to its high observing cadence. While the majority of sources were discovered via CHIMEMCA, PSR J1943+5815 and PSR J1130+0921 were noticed serendipitously by CHIME/FRB operators. We list all details of our pulsars/RRATs in Table \ref{tab:all_psr}.

\subsection{RRATs}\label{sec:rrats} 
\begin{figure*}
\centering
\includegraphics[width=\linewidth]{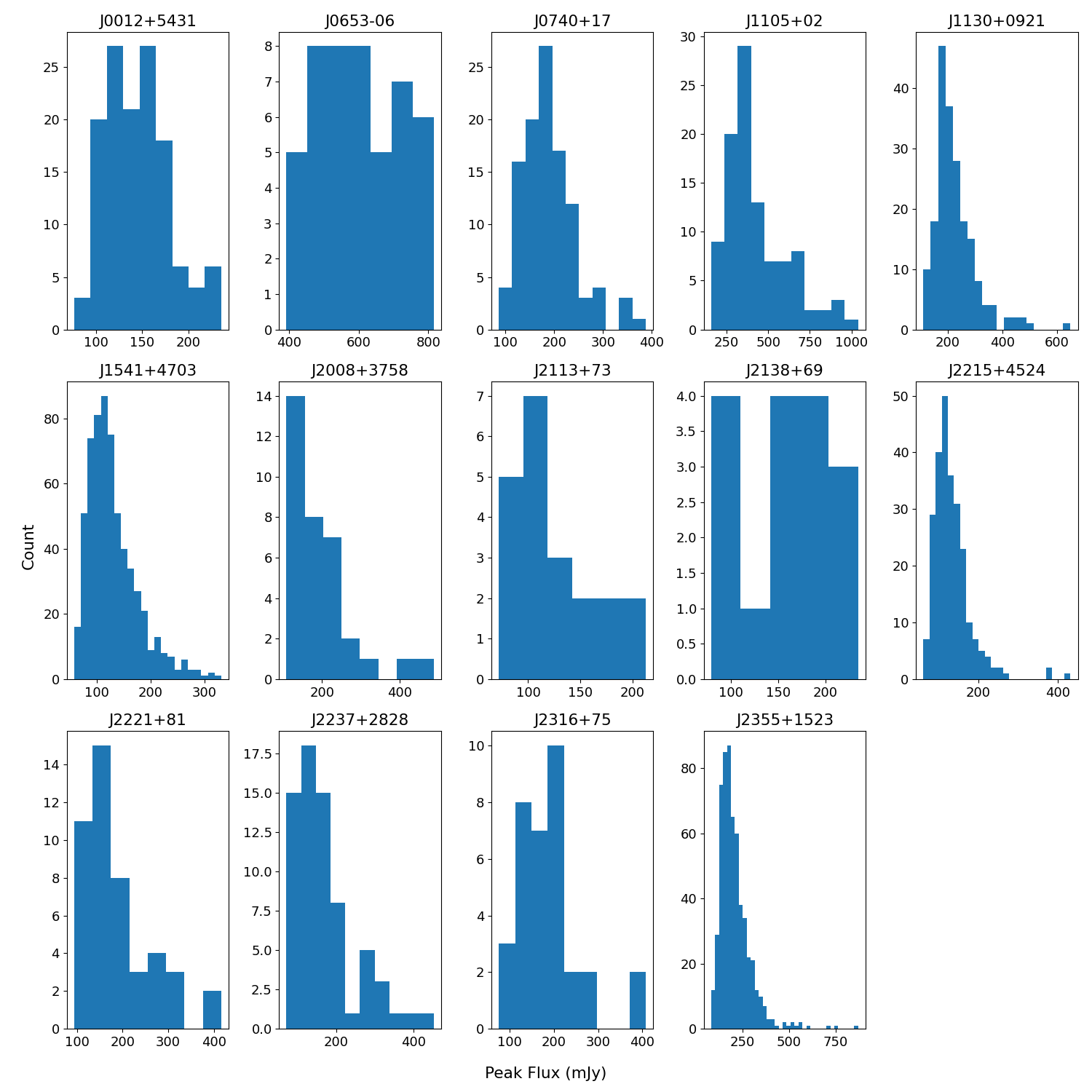}
\caption{Distributions of all the peak fluxes for the detected RRAT pulses. }
\label{fig:peak_flux}
\end{figure*}

\begin{figure*}
\centering
\includegraphics[width=\linewidth]{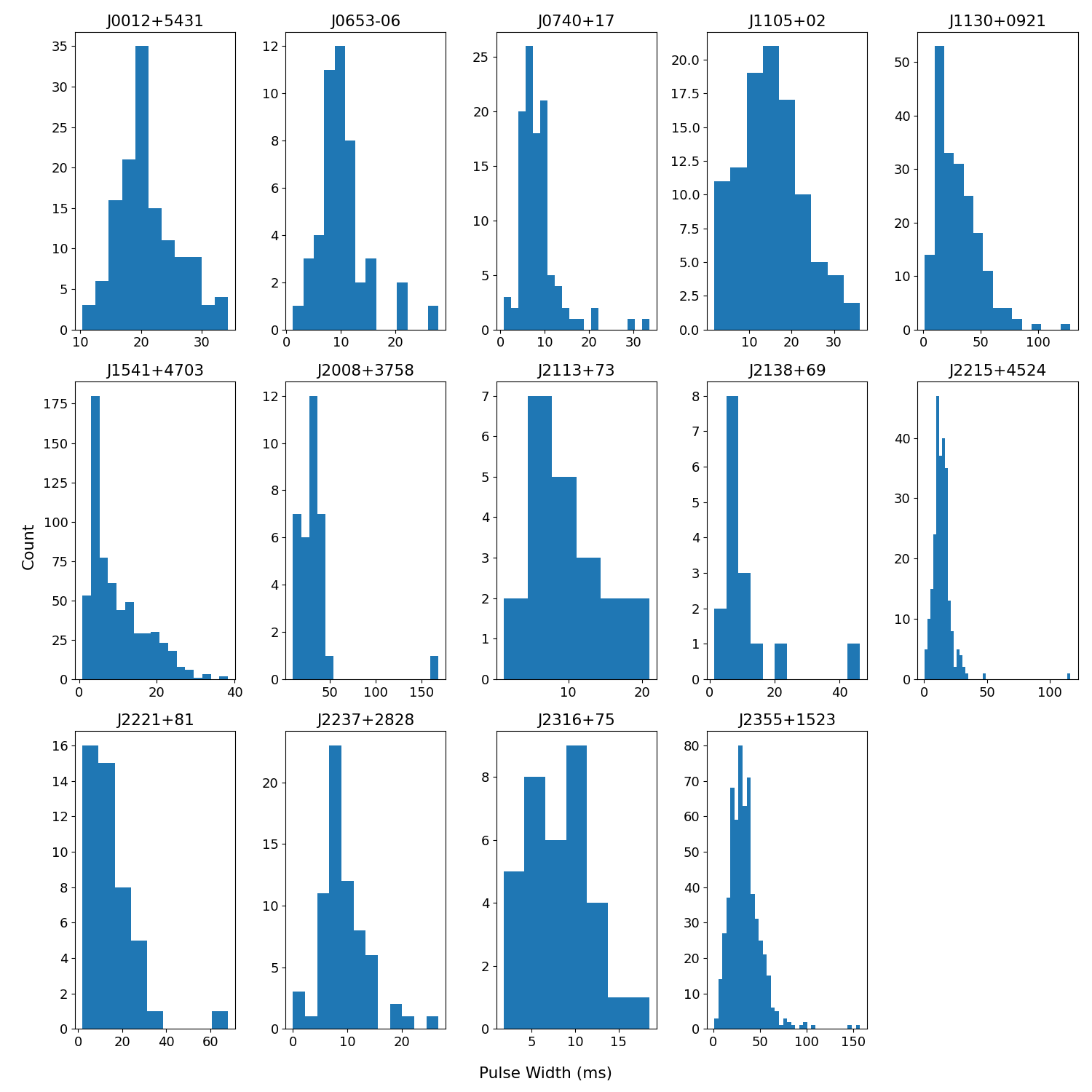}
\caption{Distributions of burst widths for the RRATs in the sample. The widths provided are the FWHM of the pulses, assuming a gaussian burst profile.}
\label{fig:widths}
\end{figure*}

\begin{figure*}
\includegraphics[width=0.8\linewidth]{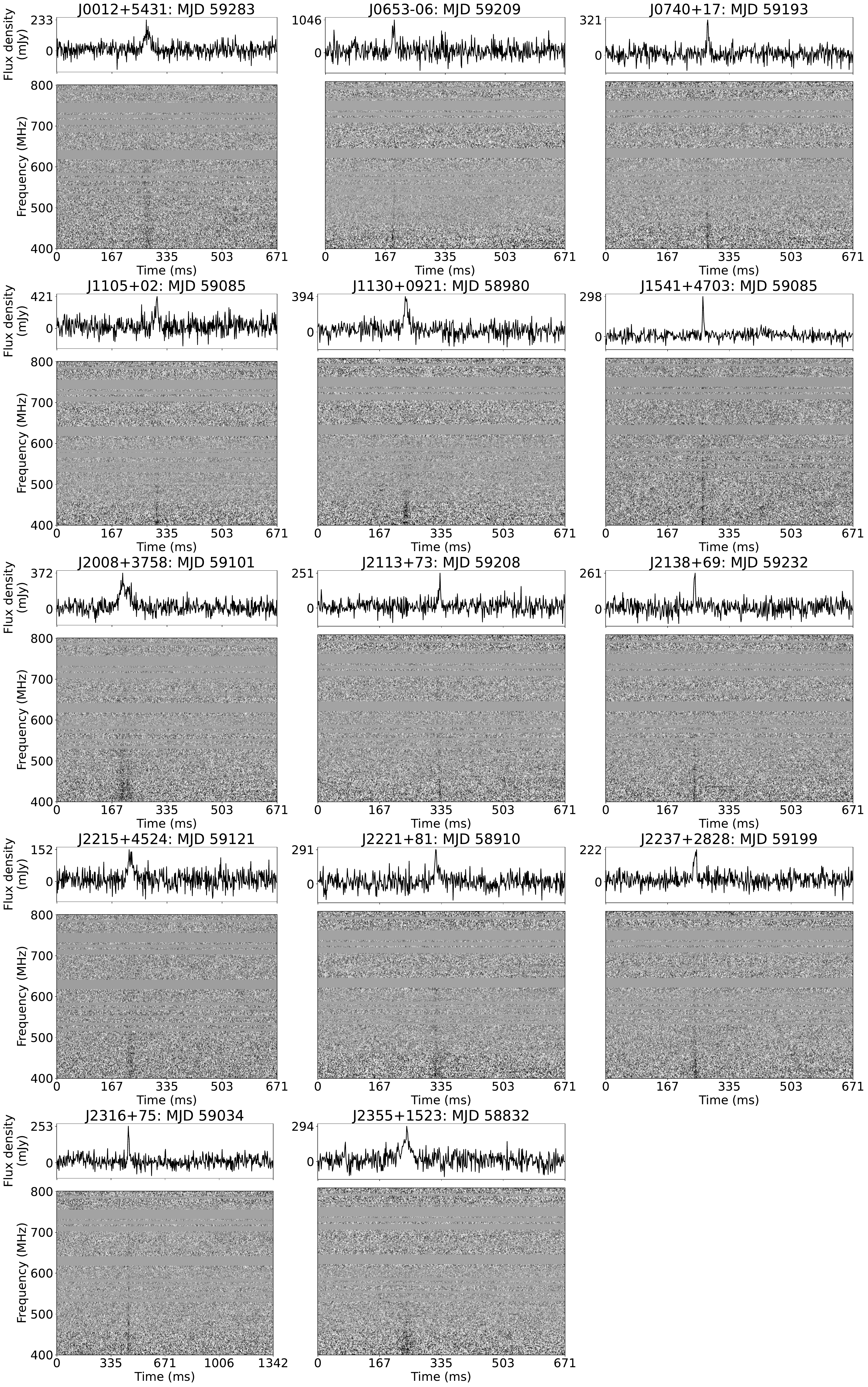}
\caption{Waterfall plots of all RRATs discovered in this study. The RFI bands have been replaced with the median and the sources have been flux calibrated.}
\label{fig:rrat_waterfalls}
\end{figure*}

In this section, we detail specific features of particular RRATs. Along with detailed texts for the most interesting RRATs, we provide Figures \ref{fig:peak_flux} and \ref{fig:widths} for the detected single pulse distributions for peak fluxes and widths respectively. We note that as in \citet{Good_2021}, the flux estimates we provide should only be treated as lower limits. The peak flux and the width FWHM was found by a maximum likelihood fit of a Gaussian to the pulse. Furthermore, we show example waterfalls for each source in \Cref{fig:rrat_waterfalls}, where each source is downsampled in both frequency and time by a factor of 4, resulting in 256 frequency channels and a time resolution of $1.31 \text{ms}$. Finally, the detection statistics for each RRAT are given in Table \ref{tab:rrat_det_stats}, timing solutions are given in Table \ref{tab:rrat_timing} and localisations are given in Table \ref{tab:rrat_pos}. Note that the localisations given in Table \ref{tab:rrat_pos} are obtained via metadata localisation of CHIME/FRB data \citep{catalog2021} and are only provided for the cases where timing localisations are not available or only partially available.\\

In \Cref{fig:peak_flux} we plot the peak flux of single pulses in the sample of RRATs. The peak fluxes generally seem to follow a log-normal like distribution. This has been well documented in other analyses of RRAT pulses such as \citep[e.g.]{Cui2017}, with others finding that fluence distributions are also distributed log-normally \citep{Burke-Spolaor2011,Meyers2018,Meyers2019,Mickaliger2018}. Without performing corrections for the selection effects of CHIPSPIPE, it is difficult to disentangle the degeneracy between the RRAT single pulse distribution and the low SNR/flux cutoff for the detection pipeline. Initial analysis shows that CHIPSPIPE is fairly complete to the SNR threshold in this study for narrow pulses with simple profiles. However the ability to recover pulses drops as they get wider and/or exhibit complex structures. Thus, we defer this analysis to future publications and note that here, we only present the characteristics of detected single pulses.\\

The width distribution of the single pulses, shown in \Cref{fig:widths}, was found by applying a maximum likelihood fit to each pulse with a single component gaussian. Then we found the full width half maxima using $\text{FWHM}=2.355\sigma$. In this analysis, we do not correct for complex structure (i.e. like those found for PSR J1105+02, PSR J1130+0921, PSR J1541+4703, PSR J2008+3758, PSR J2221+81, PSR J2355+1523) however, we have removed all the complex pulses from the histograms. This causes pulses to appear wider as the single gaussian will try to envelop multiple peaks. Furthermore, the widths of pulses are not consistent across the whole 400 MHz band; here we only provide the widths of the band-averaged profiles. Lastly, the width analysis suffers from similar shortcomings as the peak flux distributions in that we have not corrected for the selection effects of CHIPSPIPE. Nevertheless, we present the pulse-width distributions of all detected pulses and note that comparisons are still useful with this sample, as similar selection biases affect all 14 RRATs. We find that there exists large scatter in the detected pulse widths ranging from $<$0.025$\%$ (PSR J1541+4703, PSR J2008+3758, PSR J2221+81, PSR J2237+2828, PSR J2355+1523) to 13.7$\%$ (PSR J2355+1523) of the pulse period.\\

\begin{table*}
    \centering
    \caption{Detection statistics for RRATs presented in this study. Included are both the CHIME/FRB and CHIME/Pulsar detection counts. The first and last detections are given in MJD.}
    \begin{tabular}{|l|c|c|c|c|c|c|c|c|}
        \hline
        RRAT Name & FRB first det & FRB last det & $N_{det}$ FRB & Pulsar first det & Pulsar last det & $N_{det}$ Pulsar & $N_{det}$ total& \\
        \hline\hline
        PSR J0012+5431       &59370&58528&549& 58901 & 59358 &129&678&  \\
        PSR J0653-06       &59412&59492&114& 59191 & 59357 & 47& 161 &\\
        PSR J0741+17       &59381&59532&118& 59191 & 59357 &107& 225 &\\
        PSR J1105+02       &59447&59526&64& 59075 & 59359 &112& 176  &\\
        PSR J1130+0921       &59572&59378&8& 58979 & 59386 &208& 216   &\\
        PSR J1541+4703       &59371&59530&184& 59076 & 59359 &639&  823&\\
        PSR J2008+3758       &59405&59532&233& 59078 & 59346 &44& 277&\\
        PSR J2113+73       &58453&59480&26& 59203 & 59339 &21&  47   &\\
        PSR J2138+69       &58444&59441&34& 59332 & 59321 &16&  50   &\\
        PSR J2215+4524       &58401&59457&69& 59078 & 59532 &247& 316&\\
        PSR J2221+81       &58443&59495&17& 58909 & 59359 &52&  69   &\\
        PSR J2237+2828       &58424&59525&56& 59193 & 59533 &69&  125&\\
        PSR J2316+75       &58388&59491&46& 58968 & 59073 &35&  81   &\\
        PSR J2355+1523       &58584&59497&57& 58663 & 59358 &614& 871&\\\hline
    \end{tabular}
    \label{tab:rrat_det_stats}
\end{table*}

\begin{table*}
\caption{Timing solutions for 8 RRATs. A lack of errors for some positions indicates that there were insufficient TOAs to fit them. For sources without (or with only partial) positional fits, the metadata localisations and uncertainties are provided in Table \ref{tab:rrat_pos}. In these cases, the position in our timing models do not indicate the true position of the source.\\
\textdagger{} denotes a 2 $\sigma$ lower/upper limit for $\dot{\nu}$/$\dot{P}$ respectively.}
\centering
\footnotesize
\begin{tabular}{lllll}
\hline Pulsar name                                                           & PSR J0012+5431         & PSR J1105+02         & PSR J1130+0921        & PSR J1541+4703         \\
\hline Right ascension, J2000 (h:m:s)\dotfill                                & 00:12:23.3(1)    & 11:05:32.0(2)    & 11:30:55.0(5)   & 15:41:05.54(2) \\
Declination, J2000 (d:m:s)\dotfill                                           & +54:31:40(9)     & 02:28:50      & +09:21:09(14) & +47:03:03.7(3)  \\
Pulse frequency, $\nu$ (Hz)\dotfill                                          & 0.33054565343(2) & 0.15617544155(9) & 0.2084794003(3) & 3.60099929741(4) \\
Pulse frequency derivative, $\dot{\nu}$ ($\rm 10^{-16}\ Hz\,s^{-1}$)\dotfill & -0.14 \textsuperscript{\textdagger{}}           & -0.35   \textsuperscript{\textdagger{}}         & 1.2(2)          & -27.2(1)         \\
Period Epoch (MJD)\dotfill                                                   & 59128            & 59216            & 59180           & 59211            \\
Dispersion measure, DM ($\rm pc\,cm^{-3}$)\dotfill                           & 131.3(7)            & 16.5(4)           & 21.0(9)              & 19.4(7)             \\
Clock standard\dotfill                                                       & TT(BIPM2019)     & TT(BIPM2019)     & TT(BIPM2019)    & TT(BIPM2019)     \\
Ephemeris\dotfill                                                            & DE440            & DE440            & DE440           & DE440            \\
Number of TOAs\dotfill                                                       & 104              & 74               & 50              & 77               \\
Reduced $\chi^2$\dotfill                                                     & 0.82             & 0.8              & 0.953           & 1.3474           \\
\hline \multicolumn{4}{c}{Derived parameters} \\ \hline                      &                  &                  &                 &                  \\
Pulse period, $P$ (s)\dotfill                                                & 3.0253007099     & 6.403055372      & 4.796636974(6)     & 0.277700692893(3)        \\
Pulse period derivative, $\dot{P}$ ($\rm 10^{-15}\ s\,s^{-1}$)\dotfill       & 1.3  \textsuperscript{\textdagger{}}            & 14.3   \textsuperscript{\textdagger{}}          & 2.9(5)             & 0.2102(9)             \\
Surface magnetic field, $B$ ($\rm 10^{11}\,G$)\dotfill                       & -                & -                & 37.50           & 2.44             \\
Characteristic age, $\tau_c$ (Myr)\dotfill                                   & -                & -                & 26.44           & 20.87            \\
\hline                                                                       &                  &                  &                 &                  \\
\hline Pulsar name                                                           & PSR J2008+3758         & PSR J2215+4524         & PSR J2237+2828        & PSR J2355+1523         \\
\hline Right ascension, J2000 (h:m:s)\dotfill                                & 20:07:58.7(2)    & 22:15:46.57(7)   & 22:37:29.41(4)  & 23:55:48.62(8)   \\
Declination, J2000 (d:m:s)\dotfill                                           & +37:58:13(1)     & +45:24:44(2)     & +28:28:40(4)  & +15:23:19(2)     \\
Pulse frequency, $\nu$ (Hz)\dotfill                                          & 0.2297634166(1)  & 0.36723529308(4) & 0.9281646147(2) & 0.91374580877(9) \\
Pulse frequency derivative, $\dot{\nu}$ ($\rm 10^{-16}\ Hz\,s^{-1}$)\dotfill & -0.38  \textsuperscript{\textdagger{}}          & -7.5(3)          & -7(2)           & -3.5(2)          \\
Period Epoch (MJD)\dotfill                                                   & 59212            & 59241            & 59289           & 59121            \\
Dispersion measure, DM ($\rm pc\,cm^{-3}$)\dotfill                           & 143(1)        & 18.5(4)             & 38.1(4)            & 26(1)           \\
Clock standard\dotfill                                                       & TT(BIPM2019)     & TT(BIPM2019)     & TT(BIPM2019)    & TT(BIPM2019)     \\
Ephemeris\dotfill                                                            & DE440            & DE440            & DE440           & DE440            \\
Number of TOAs\dotfill                                                       & 39               & 103              & 22              & 128              \\
Reduced $\chi^2$\dotfill                                                     & 2.2              & 0.87             & 1.97            & 1.17             \\
\hline \multicolumn{4}{c}{Derived parameters} \\ \hline                      &                  &                  &                 &                  \\
Pulse period, $P$ (s)\dotfill                                                & 4.352302968      & 2.7230498235(3)     & 1.0773950914(7)    & 1.09439626467(5)    \\
Pulse period derivative, $\dot{P}$ ($\rm 10^{-15}\ s\,s^{-1}$)\dotfill       & 7.2    \textsuperscript{\textdagger{}}          & 5.6(2)              & 0.9(3)            & 0.41(2)             \\
Surface magnetic field, $B$ ($\rm 10^{11}\,G$)\dotfill                       & -                & 39.40            & 9.87            & 6.77             \\
Characteristic age, $\tau_c$ (Myr)\dotfill                                   & -                & 7.71             & 19.27           & 42.31            \\
\hline                                                                       &                  &                  &                 &                 
\end{tabular}
\label{tab:rrat_timing}
\end{table*}

Of the 14 RRATs in this sample, we obtained timing solutions for 8 of them. The timing models are presented in Table \ref{tab:rrat_timing}. The remaining 6 either had too few pulses, failed to have multiple pulses within one observation, or both. We produced the timing solution by extracting single pulse TOAs using the method outlined in \secref{sec:timing} and fit them to a timing model using the TEMPO2 software. In the process, the epoch was always set to the centre of the time span using the \texttt{-epoch centre} command of TEMPO2. The timing solution often lacks a $\dot{\nu}$ measurement and we hope that with continued monitoring $\dot{\nu}$ can eventually be determined.

\subsection{Complex Single Pulse Structures}
\begin{figure*}
\includegraphics[width=\linewidth]{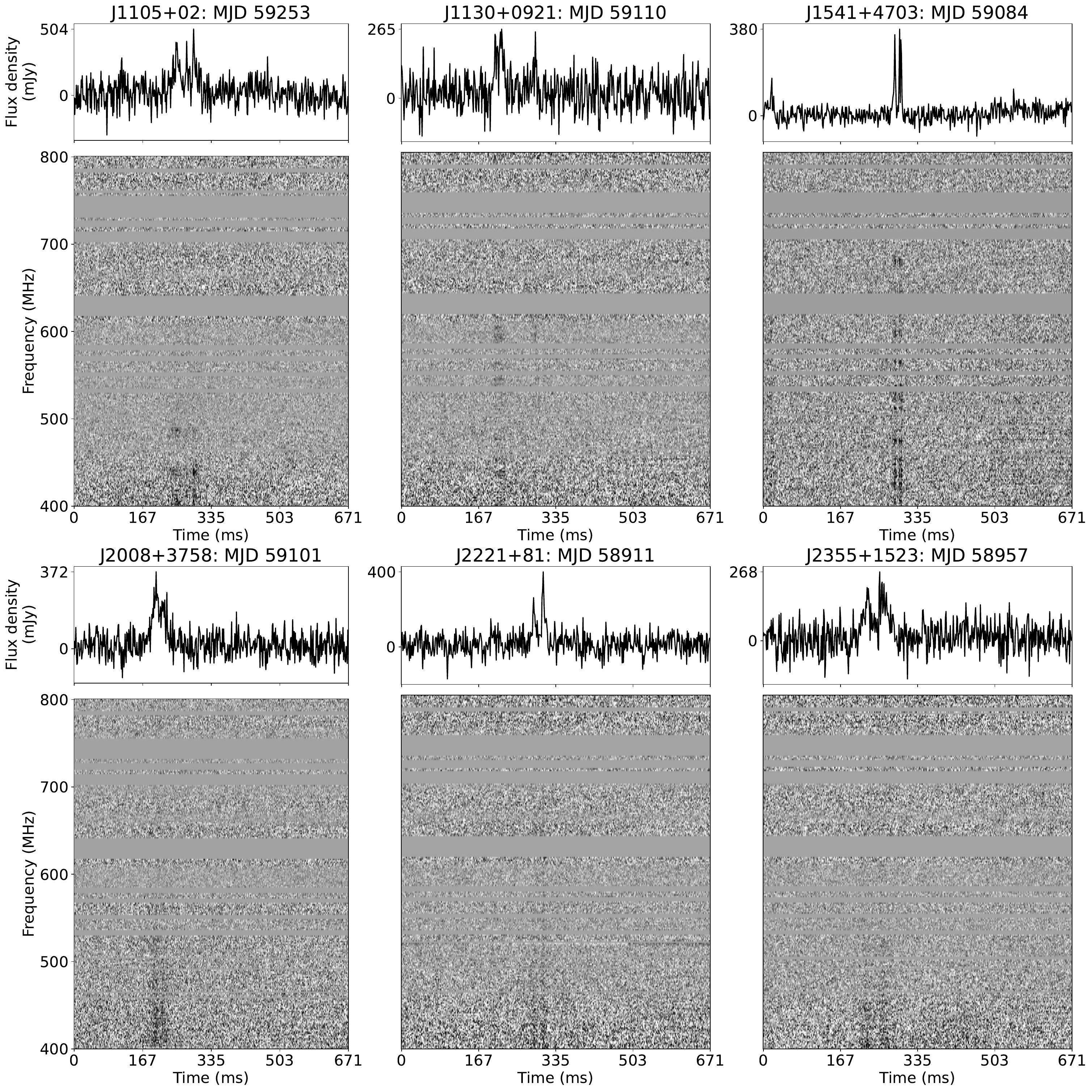}
\caption{Example waterfall plots of RRATs that show complex structure.}
\label{fig:weird_waterfalls}
\end{figure*}

Many of the RRATs in this sample exhibit a variety of single pulse features. One prolific example is double or multiple peaks; examples can be found in \Cref{fig:weird_waterfalls}. These are also often seen in other RRATs \citep{Keane2010,Bhattacharyya2018,TyulBashev2018}. In this sample, 6 of the 14 RRATs exhibit some sort of complex structure.\\
PSR J1105+02 occasionally emits inter-pulses as well as double peaks. It is unclear whether the two emission types have the same underlying mechanism. Additionally, the second peak or inter-pulse can sometimes be brighter than the initial pulse.\\
PSR J1130+0921, PSR J2008+3758, PSR J2221+81, PSR J2355+1523 are the other sources that produces double peaked pulses. Two RRATs, PSR J1130+0921 and PSR J1541+4703, also produce complex single pulse structure but are discussed separately below.\\

\subsubsection{PSR J1105+02}
PSR J1105+02 is a bright RRAT detected by both CHIME/FRB and CHIME/Pulsar. As we processed the data, we found that there was a frequency dependent structure that indicated a slightly erroneous pointing, based on the known CHIME beam shape. We have corrected the position with a single pulse grid search whereby six beams are placed simultaneous on the sky. The best-fit positions are further corroborated by a timing analysis. Thus, the source is probably much brighter than what we have shown in \Cref{fig:peak_flux}. The source is among the brightest in the sample as we still detect peak fluxes in excess of 1\,Jy even with the position offset. However, we should note that due to the position offset, the peak fluxes shown in \Cref{fig:peak_flux} likely have greater uncertainties than for all other sources.\\

\subsubsection{PSR J1541+4703}
CHIME/Pulsar collected 248 search-mode observations for a total of 79.05 hours on PSR J1541+4703. The initial detection was on MJD 59076. Subsequently, CHIPSPIPE found 639 bursts making PSR J1541+4703 the most prolific RRAT in the sample. Among these bursts we find some spectral variation, sub-pulse drifting and extremely narrow pulses that are $<$0.025\% of the pulse period. Finally, this RRAT also often displays double peaks.\\
\begin{table*}
\caption{Localisation of sources without robust timing solutions. The localisations are provided via the metadata localisation method \protect\citep{catalog2021}. The metadata localisation method can often result in multiple regions, here we give only the peak of the probability distribution and quote the 95\% confidence interval. Missing periods indicate insufficient days where multiple pulses were detected. Sources with only partial timing localisations are included here also with the lowest uncertainty positions listed.}
\centering
\footnotesize
\begin{tabular}{lllll}
\hline
PSR Name & RA(h:m:s) & Dec (d:m:s) & DM (pc/cm$^3$)& Period (s)\\ \hline\hline
PSR J0653-06 & 6:53(1)  & -6:16(36) & 83.7 & 0.79 \\
PSR J0741+17 & 7:41(1)  & 17:03(23)   & 44.3 & 1.73 \\
PSR J1105+02 & 11:05:32.0(2)  & 2:28(30)   & 16.5 & 6.4  \\
PSR J2113+73 & 21:13(1) & 73:37(21)  & 42.4 & --   \\
PSR J2138+69 & 21:38(1) & 69:50(16)  & 46.6 & 0.22 \\
PSR J2316+75 & 23:15(1) & 75:44(21)  & 53.4 & --   \\
PSR J2221+81 & 22:21(1) & 81:32(30)  & 39   & --   \\ \hline
\end{tabular}
\label{tab:rrat_pos}
\end{table*}

\subsection{Regular pulsars}
All 7 regular pulsars in the sample are so called slow pulsars with periods >100ms. Their timing solutions are given in Table \ref{tab:pulsar_timing}. For most sources, it was immediately obvious that the discovery was of a regularly pulsating source, often confirmed by the \texttt{prepfold} software from \texttt{PRESTO} or a fast folding algorithm like \texttt{riptide} \citep{Morello2020}. These sources were quickly switched to fold-mode observations and their timing models were found via TEMPO2. PSR J1048+5349 was the only exception where we did not realise that the source was a regular pulsar, thus TOAs for PSR J1048+5349 were derived from folded search-mode observations.
\subsubsection{PSR J2208+4610}
PSR J2208+4610 was initially detected as a regular pulsar with a period of 0.642\,s and a DM of $\sim63$ $\text{pc\,cm}^{-3}$. Subsequent observations of the pulsar showed variation in rotation properties that required modelling beyond the first spin frequency derivative. Furthermore, the fold-mode data obtained from the pulsar showed signs of varying spin period over time, suggesting that the pulsar is in a binary system. Further observations revealed that the spin variation of PSR J2208+4610 is consistent with it being in a binary system with a period of roughly 412 days and a low eccentricity of 0.03. The timing solution of the pulsar is presented in Table~\ref{tab:J2206}. The binary companion of PSR J2208+4610 is found to have a minimum mass of 0.1303 solar masses, assuming a pulsar mass of 1.4 solar masses. This suggests that the companion is most likely a Helium white dwarf. Such binary systems are rare, with only a few of them known, e.g. PSR B0820+02~\citep{Koester2000} and PSR B1800-27~\citep{Johnston1995}, as most pulsar-Helium white dwarf systems consist of a millisecond pulsar of period $<10$ ms in a 
circular orbit. A search in archival imaging surveys data revealed no visible optical companion with 10'' of the pulsar position. This is not surprising considering the DM-distance estimate of the pulsar is $\sim3$ kpc~\citep{Yao2016}.

\begin{table}
\caption{The timing solution for J2208+4610}
\centering
\footnotesize
\begin{tabular}{ll}
\hline Pulsar name                                                           & PSR J2208+4610        \\
\hline Right ascension, J2000 (h:m:s)\dotfill                                & 22:08:23.36(2) \\
Declination, J2000 (d:m:s)\dotfill                                           & +46:10:04.96(4)  \\
Pulse frequency, $\nu$ (Hz)\dotfill                                          & 1.5564067924676(12) \\
Pulse frequency derivative, $\dot{\nu}$ ($\rm 10^{-16}\ Hz\,s^{-1}$)\dotfill & -0.065(3)         \\
Binary Model \dotfill                                                        & BT\\
Orbital Period, $P_{\rm b}$ (Days)\dotfill                                   & 412.4866(4)\\
Projected semi-major axis, $x$ (lt-s)\dotfill                                & 54.2885(4)\\
Epoch of periastron passage, $T_0$ (MJD)\dotfill                             & 59221.6789(16)\\
Longitude of periastron, $\omega$ (deg) \dotfill                             & 358.7179(14)\\
Orbital eccentricity, $e$\dotfill                                            & 0.0328479(8)\\
Period Epoch (MJD)\dotfill                                                   & 59540            \\
Dispersion measure, DM ($\rm pc\,cm^{-3}$)\dotfill                           & 63.407             \\
Clock standard\dotfill                                                       & TT(BIPM2019)\\
Ephemeris\dotfill                                                            & DE440\\
Number of TOAs\dotfill                                                       & 325\\
Reduced $\chi^2$\dotfill                                                     & 1.13          \\
\hline \multicolumn{2}{l}{Derived parameters} \\ \hline                      &\\
Pulse period, $P$ (s)\dotfill                                                & 0.6425055485748(5)       \\
Pulse period derivative, $\dot{P}$ ($\rm 10^{-15}\ s\,s^{-1}$)\dotfill       & 0.00268(11) \\
Surface magnetic field, $B$ ($\rm 10^{11}\,G$)\dotfill                       & 0.42044             \\
Characteristic age, $\tau_c$ (Myr)\dotfill                                   & 3791.6            \\
Mass function\dotfill                                                        & 0.00100969(2)\\
Minimum companion mass\dotfill                                               & 0.1303$M_\odot$\\
\hline
\end{tabular}
\label{tab:J2206}
\end{table}

\begin{table*}
\caption{The timing solutions for discovered regular pulsars apart from J2208+4610}
\centering
\footnotesize
\begin{tabular}{llll}
\hline Pulsar name                                                           & PSR J0227+3356          & PSR J0658+2936         & PSR J1048+5349         \\
\hline Right ascension, J2000 (h:m:s)\dotfill                                & 02:26:57.060(1)  & 06:57:44.62(8)  & 10:48:06.46(5)  \\
Declination, J2000 (d:m:s)\dotfill                                           & +33:56:27.76(7)  & +29:36:29(7)    & +53:49:51.9(3)  \\
Pulse frequency, $\nu$ (Hz)\dotfill                                          & 0.806384981978(2) & 1.2140327894(6)  & 0.36618479871(2) \\
Pulse frequency derivative, $\dot{\nu}$ ($\rm 10^{-16}\ Hz\,s^{-1}$)\dotfill & -18.542(7)        & -21.2(6)         & -1.63(9)         \\
Period Epoch (MJD)\dotfill                                                   & 59304             & 59266            & 59216            \\
Dispersion measure, DM ($\rm pc\,cm^{-3}$)\dotfill                           & 27(1)             & 39(1)            & 27.3(9)           \\
Clock standard\dotfill                                                       & TT(BIPM2019)      & TT(BIPM2019)     & TT(BIPM2019)     \\
Ephemeris\dotfill                                                            & DE440             & DE440            & DE440            \\
Number of TOAs\dotfill                                                       & 314               & 141              & 215              \\
Reduced $\chi^2$\dotfill                                                     & 1.3415            & 1.6144           & 1.237            \\
\hline \multicolumn{4}{c}{Derived parameters} \\ \hline                      &                   &                  &                  \\
Pulse period, $P$ (s)\dotfill                                                & 1.2401024601760(3)    & 0.8237009813(4)     & 2.7308615854955(11)  \\
Pulse period derivative, $\dot{P}$ ($\rm 10^{-15}\ s\,s^{-1}$)\dotfill       & 2.8515(11)             & 1.44(4)             & 1.18(8)             \\
Surface magnetic field, $B$ ($\rm 10^{11}\,G$)\dotfill                       & 19.03             & 11.02            & 18.46            \\
Characteristic age, $\tau_c$ (Myr)\dotfill                                   & 6.87              & 9.04             & 35.39            \\
\hline Pulsar name                                                           & PSR J1943+5815          & PSR J2057+4701         & PSR J2116+3701         \\
\hline Right ascension, J2000 (h:m:s)\dotfill                                & 19:43:23.175(2)  & 20:57:09.335(2) & 21:16:14.304(1) \\
Declination, J2000 (d:m:s)\dotfill                                           & +58:15:53.08(2)  & +47:01:05.22(3) & +37:01:31.13(4) \\
Pulse frequency, $\nu$ (Hz)\dotfill                                          & 0.786876829081(1) & 1.78707496869(2) & 6.85475255154(2) \\
Pulse frequency derivative, $\dot{\nu}$ ($\rm 10^{-16}\ Hz\,s^{-1}$)\dotfill & -2.534(4)         & -1789.66(1)      & -2475(1)         \\
Period Epoch (MJD)\dotfill                                                   & 59465             & 59249            & 59219            \\
Dispersion measure, DM ($\rm pc\,cm^{-3}$)\dotfill                           & 71(1)             & 219.0(5)         & 43.7(2)               \\
Clock standard\dotfill                                                       & TT(BIPM2019)      & TT(BIPM2019)     & TT(BIPM2019)     \\
Ephemeris\dotfill                                                            & DE440             & DE440            & DE440            \\
Number of TOAs\dotfill                                                       & 43                & 187              & 170              \\
Reduced $\chi^2$\dotfill                                                     & 0.9487            & 1.1783           & 1.545            \\
\hline \multicolumn{4}{c}{Derived parameters} \\ \hline                      &                   &                  &                  \\
Pulse period, $P$ (s)\dotfill                                                & 1.270846926148(4)     & 0.559573614715(7)   & 0.1458841865523(4)  \\
Pulse period derivative, $\dot{P}$ ($\rm 10^{-15}\ s\,s^{-1}$)\dotfill       & 4.073(11)             & 56.0382(3)          & 5.2674(3)           \\
Surface magnetic field, $B$ ($\rm 10^{11}\,G$)\dotfill                       & 7.30              & 56.67            & 8.87             \\
Characteristic age, $\tau_c$ (Myr)\dotfill                                   & 49.06             & 0.16             & 0.44  \\
\hline
\end{tabular}
\label{tab:pulsar_timing}
\end{table*}

\section{discussion of results}\label{sec:discussion}
\begin{figure*}
\centering
\includegraphics[width=0.85\linewidth]{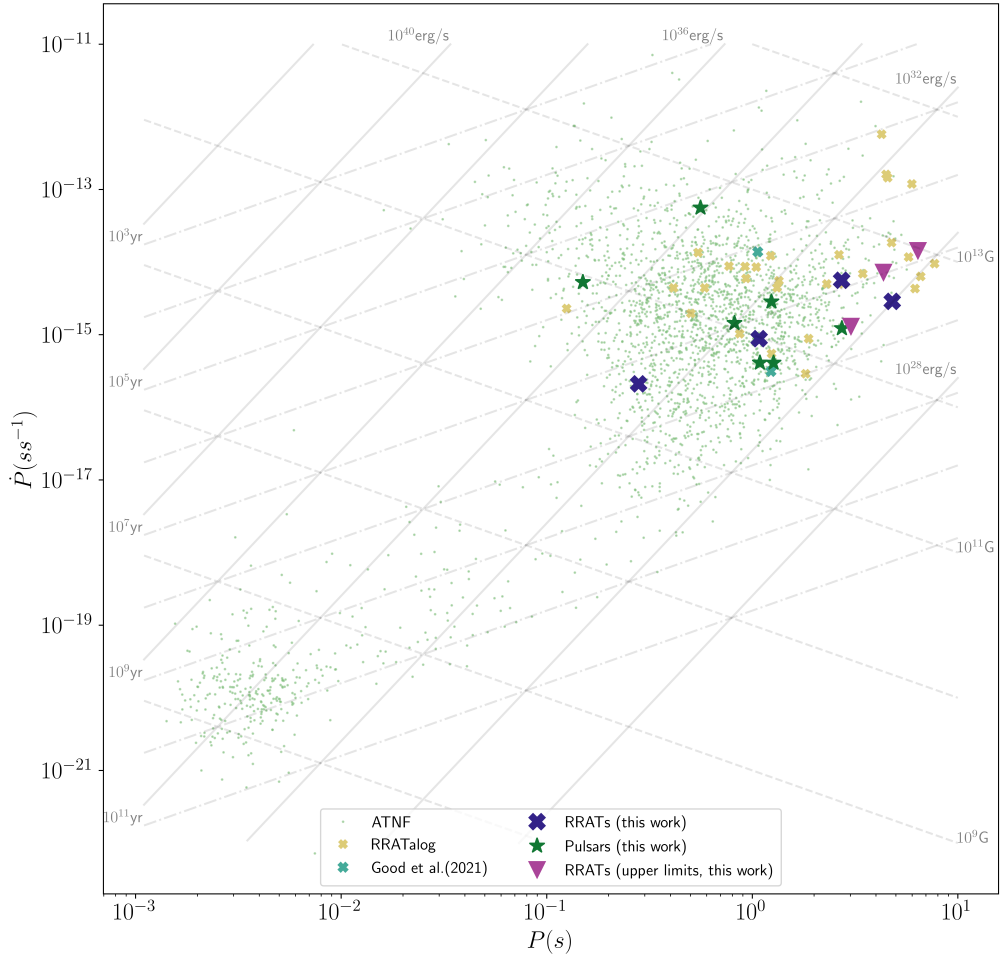}
\caption{$P-\dot{P}$ diagram that includes all pulsars on the ATNF catalog (version 1.7, magnetars and XDINs removed) and RRATs in the RRATalog (as of August 2022). Discoveries made by CHIME/FRB in both \protect\cite{Good_2021} and this work have been overlaid. For 3 RRATs we plot only upper limits for the measurement on $\dot{P}$. Only RRATs/pulsars with P and $\dot{\text{P}}$ measurements are plotted in this figure. Lines of constant magnetic field, characteristic age and spin down luminosity are denoted by dashed and solid lines with the appropriate units.}
\label{fig:ppdot}
\end{figure*}
Throughout this work and together with that reported by \citet{Good_2021} we have shown that CHIME/FRB is an good tool for finding new pulsars and RRATs. Furthermore, CHIME/FRB’s daily cadence of observations of the northern sky and exceptional exposure have poised it for population studies of RRATs. Employing unsupervised machine learning we have found an order of magnitude more pulsars than relying on serendipitous human discoveries. Other wide field-of-view telescopes (i.e. CRAFT, LOFAR) could recruit CHIMEMCA-like analyses to aid their searches.\\
We have placed the RRATs and pulsars in a $P-\dot{P}$ diagram in \Cref{fig:ppdot}; all the discovered sources lie in the region of long-period pulsars. Among the discovered sources, most were simple-peaked isolated pulsars, however a few of the RRATs displayed interesting single pulse properties such as double peaks and spectral variation. This could motivate further investigation for RRATs such as PSR J1541+4703. Furthermore, one source is a detached binary with an unusually long period, PSR J2208+4610, highlighting CHIME/FRB/Pulsar potential for discoveries of interesting sources.\\
As the CHIME/FRB survey progresses, we will begin to map a large portion of the RRATs above CHIME/FRB's sensitivity threshold in the northern hemisphere. Due to CHIME/FRB's exposure time, we have an opportunity to perform detailed statistics on the RRAT population and their distributions that has been difficult to achieve using single-dished telescopes.

\subsection{Comparison of new CHIME sources to the known RRAT population}
We compare the properties of the newly discovered RRATs to that of the broader population using the same method as outlined by \citet{Good_2021}, and again note that the publicly accessible information about historical RRAT discoveries in surveys (or data reprocessing) is, for the most part, not complete in the sense that selection/observation biases are unchecked.

In this sample, the predominant selection biases would relate to the burst rates and pulse widths. 
Many of the known RRATs were discovered with single-dish telescopes through surveys that have dwell times of a few minutes at maximum, which limits the detected RRATs to be those with burst rates (or wait-time distributions) that correspond to those time scales or shorter. 
In the context of burst widths, CHIME/FRB samples at 1-ms time intervals whereas typical pulsar surveys use $\sim$10--100\,$\mu$s time resolution, meaning that faint and narrow bursts that would otherwise be detectable in the pulsar survey data would appear with reduced in S/N when observed through the CHIME/FRB system. 
This reduction in S/N could be significant enough, depending on the burst structure, that it falls below the online CHIME/FRB detection thresholds and, therefore, would never be followed up with the CHIME/Pulsar system which could resolve such ``narrow'' bursts. 
This width bias is further exacerbated by the apparent reduced completeness for wide/complex bursts in CHIPSPIPE, as noted in Section~\ref{sec:rrats}, which limits the recovery of wider bursts, i.e., our sample is ``windowed'' in the pulse width parameter space. 
Due to these short-falls, we emphasise that the results presented here are purely observational.
Rigorously exploring the astrophysical importance would require knowing the biases for each survey and correcting the samples on a case-by-case basis, at which point we would also be dealing with small-number statistics. 
For that reason, we take this analysis as a naive rationality check.

\begin{figure*}
    \centering
    \includegraphics[width=0.85\textwidth]{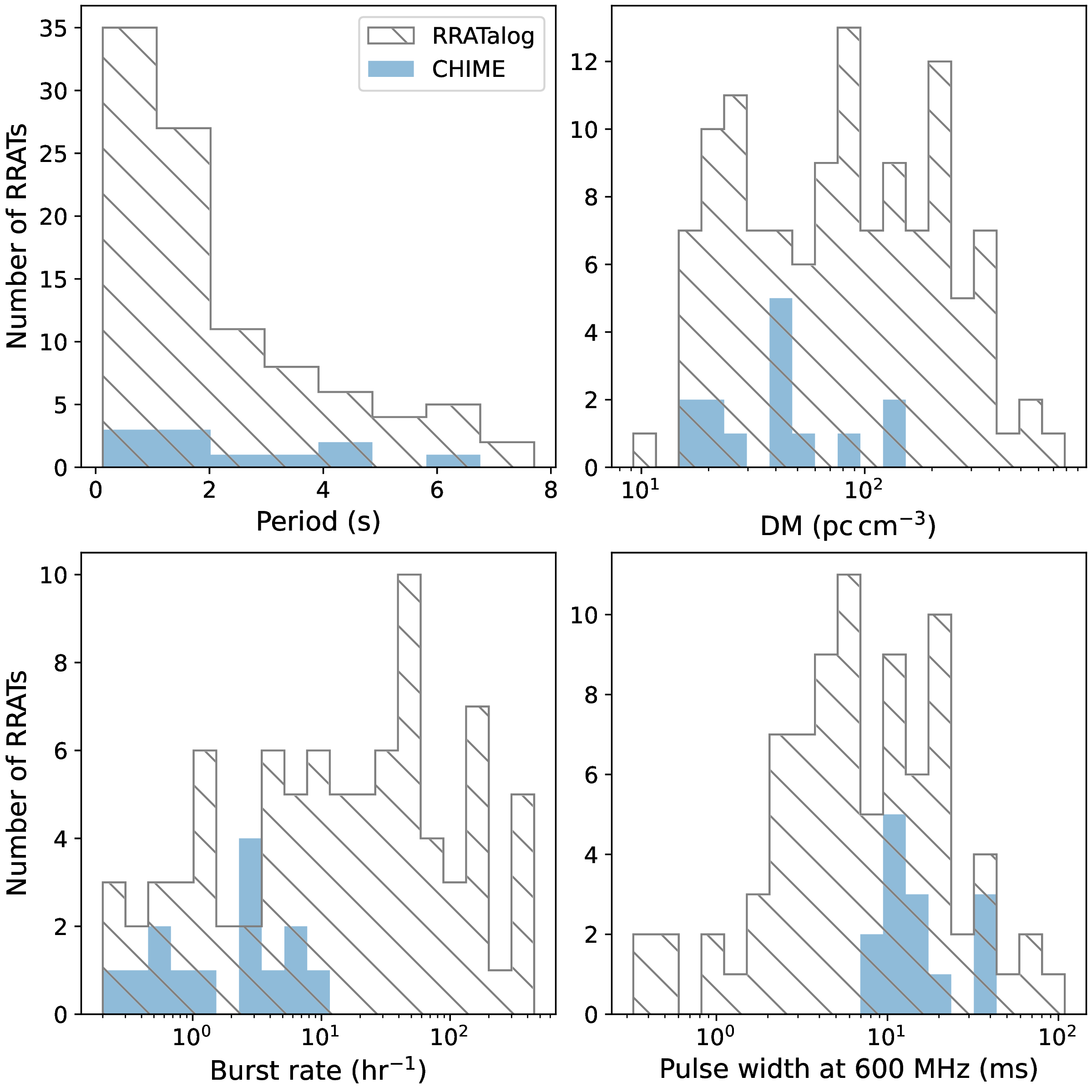}
    \caption{Rotation period (top left), dispersion measure (top right), burst rate (bottom left), and pulse width (bottom right) distributions of known RRATs (grey-hatched histogram) and the CHIME candidates presented here (blue histogram). For the width distributions, we have scaled all widths to 600\,MHz following the procedure of \citet{Good_2021}. Distribution sizes and a summary of their statistical comparison are given in Table~\ref{tab:rrat_pop_comp}.}
    \label{fig:rrat_pop_comp}
\end{figure*}

The period, DM, burst rate, and pulse width distributions for both known RRATs and CHIME sources are shown in \Cref{fig:rrat_pop_comp}.
We include only those sources presented in Section~\ref{sec:rrats}.
Given the relatively small sample size, we bootstrap the Kruskal-Willis H-test\footnote{The null hypothesis for this test is, essentially, that the two samples are drawn from a distribution with the same median. We use Python's `scipy.stats.kruskal` implementation.} statistic computation, resampling both the CHIME and known RRAT samples with replacement and iterating $10^4$ times for each distribution.

The 95-percentile range of p-values from these tests (see Table~\ref{tab:rrat_pop_comp}) are then used to decide whether the null hypothesis can be rejected -- we reject the null hypothesis only if the upper confidence bounds are below 0.05. We note that, due to the widely varying telescope and survey configurations used to find the compared samples, the meaning of any ``significance'' is weak. 

\begin{table*}
    \centering
    \caption{Sample sizes and 95-percentile p-value ranges from the bootstrapped Kruskal-Willis H-tests, comparing the CHIME and known RRAT distributions \label{tab:rrat_pop_comp}.}
    \begin{tabular}{lcccc}
       \hline
       Parameter  & $N_{\rm CHIME}$ & $N_{\rm RRATalog}$ (As of August 2022) & \multicolumn{2}{c}{p-value range} \\
       \hline\hline
        Period      & 11 & 98  & 0.013 & 0.98 \\
        DM          & 14 & 122 & $1.9 \times 10^{-4}$ & 0.39 \\
        Burst rate  & 14 & 84  & $3.6 \times 10^{-6}$ & 0.01 \\
        Burst width & 14 & 84  & $1.6 \times 10^{-4}$ & 0.20 \\
        \hline
    \end{tabular}
\end{table*}
The only notable result is that these new candidates appear to have a slightly lower median burst rate than the broader population, although this is unsurprising given the previously outlined burst rate bias inherent within the publicly available data (e.g., CHIME/FRB and CHIME/Pulsar achieve dwell times of $\gtrsim$10 minutes per observation and can revisit the position daily, whereas other telescopes/surveys observed positions for 2--5 minutes and usually only once).
The brightest and highest burst-rate RRATs are naturally the objects discovered first by earlier large sky-area pulsar/fast-transient surveys.
We would, therefore, expect that our discoveries would tend towards being sources with much lower burst rates and/or higher burst-rate irregularities.
\section{Conclusion}
This work reports on the discovery of 21 new Galactic pulsars and RRATs. We again show that CHIME/FRB, although primarily a FRB machine, is a powerful instrument for detecting RRATs even without total intensity data. To improve the completeness of the RRAT/Pulsar survey of CHIME/FRB we have developed CHIMEMCA, a technique using the metadata collected by CHIME/FRB. Candidates identified with CHIMEMCA are followed up with CHIME/Pulsar with near-daily observations, enabling detections into the low burst-rate parameter space of RRATs. The high data rate makes processing the search-mode data too labour intensive, thus we have developed an automated pipeline, CHIPSPIPE, comprising current state-of-the-art software. With CHIMEMCA and CHIPSPIPE working synergistically, we have discovered 13 new RRATs and 6 new pulsars and 1 binary system. In addition we also report the discovery of 1 new RRAT and 1 new pulsar serendipitously.

The development of CHIPSPIPE and CHIME/Pulsar's high observation cadence opens the door for large scale studies on RRAT properties and statistics (Meyers et al. in prep). This technical development could lend itself to other telescopes hoping to carry out similar FRB and Pulsar surveys.

By obtaining timing solutions for 8 out of the 14 RRATs discovered, we show that CHIME/Pulsar is capable of timing intermittent sources. Robust timing solutions become possible even for RRATs with low burst rates. This shows great synergy with CHIME/FRB's large survey volume as pulsar timing is able to localise even low burst rate RRATs with arc second precision. We plan to continue the long term monitoring of the RRATs discovered in this work along with those discovered by \citet{Good_2021}. In the future, we will test automated methods of obtaining the timing solution for RRATs.

We discovered a binary system with an unusually long orbital period of 412 days accompanied by a low mass Helium Dwarf with a minimum mass of 0.1303 solar masses. We hope to discover many more detached binaries systems such as this as the CHIME/FRB pulsar survey progresses. We note that in this mode of searching, we are unable to perform acceleration searches from the CHIME/FRB initial detection. Yet, once search-mode data has been obtained with CHIME/Pulsar, this becomes possible. Performing an acceleration search is outside the scope of this study. However, in the future, acceleration searches will be complementary to long period timing for finding new binary pulsars.

As the CHIME/FRB experiment continues, we will pursue all the Galactic candidates that CHIMEMCA produces. In time, this will elucidate various aspects of the mysterious RRAT population. Not only will we increase the number of extant RRATs, but we will likely be able to discover a large portion of the RRATs above CHIME/FRB's sensitivity threshold in the northern hemisphere.
\section*{Acknowledgements}
We acknowledge that CHIME is located on the traditional, ancestral, and unceded territory of the Syilx/Okanagan people. We are grateful to the staff of the Dominion Radio Astrophysical Observatory, which is operated by the National Research Council of Canada.  CHIME is funded by a grant from the Canada Foundation for Innovation (CFI) 2012 Leading Edge Fund (Project 31170) and by contributions from the provinces of British Columbia, Qu\'{e}bec and Ontario. The CHIME/FRB Project, which enabled development in common with the CHIME/Pulsar instrument, is funded by a grant from the CFI 2015 Innovation Fund (Project 33213) and by contributions from the provinces of British Columbia and Qu\'{e}bec, and by the Dunlap Institute for Astronomy and Astrophysics at the University of Toronto. Additional support was provided by the Canadian Institute for Advanced Research (CIFAR), McGill University and the McGill Space Institute thanks to the Trottier Family Foundation, and the University of British Columbia. The CHIME/Pulsar instrument hardware was funded by NSERC RTI-1 grant EQPEQ 458893-2014.

This research was enabled in part by support provided by the BC Digital Research Infrastructure Group and the Digital Research Alliance of Canada (alliancecan.ca).\\
A.M.C is supported by the Ontario Graduate Scholarship\\
K.C. is supported by a UBC Four Year Fellowship (6456)\\
F.A.D is supported by a UBC Four Year Fellowship (6456)\\
JWM gratefully acknowledges support by the Natural Sciences and Engineering Research Council of Canada (NSERC), [funding reference \#CITA 490888-16]\\
B.M.G. is supported by an NSERC Discovery Grant (RGPIN-2022-03163), and by the Canada Research Chairs (CRC) program.\\
J. W. McKee is a CITA Postdoctoral Fellow: This work was supported by Ontario Research Fund—research Excellence Program (ORF-RE) and the Natural Sciences and Engineering Research Council of Canada (NSERC) [funding reference CRD 523638-18]\\
A.B.P. is a Banting Fellow, McGill Space Institute (MSI) Fellow, and a Fonds de Recherche du Quebec -- Nature et Technologies (FRQNT) postdoctoral fellow.\\
Z.P. is a Dunlap Fellow.\\
V.M.K. holds the Lorne Trottier Chair in Astrophysics \& Cosmology,a Distinguished James McGill Professorship, and receives support from an NSERC Discovery grant (RGPIN
228738-13), from an R. Howard Webster Foundation Fellowship from CIFAR, and from the FRQNT CRAQ.\\
Pulsar and FRB research at UBC are supported by an NSERC Discovery Grant and by the Canadian Institute for Advanced Research\\
%

\vspace{5mm}


Software: \texttt{PRESTO, SPEGID, FETCH, PSRCHIVE, DSPSR, TEMPO, TEMPO2}

\section{Data Availability}
The data underlying this article will be shared on reasonable request to the corresponding author.



\bibliographystyle{mnras}
\bibliography{references} 





\bsp	
\label{lastpage}
\end{document}